\documentstyle[11pt,newpasp,twoside,epsf]{article}
\begin{document}

\title{ A HEURISTIC INTRODUCTION TO RADIOASTRONOMICAL POLARIZATION} 

\author{Carl Heiles}

\affil {Astronomy Department, University of California,
    Berkeley, CA 94720-3411; cheiles@astron.berkeley.edu}

\begin{abstract}

	Radio sources are often polarized. Accurate measurement of
simply the flux density of a radio source requires a basic understanding
of polarization and its measurement techniques. We provide an
introductory, heuristic discussion of these matters with an emphasis on
practical application and avoiding pitfalls. \end{abstract}

\section{Introduction}  \label{introduction}

	Many astronomers wish nothing more than to measure the total
flux density of a source. Radio sources are often polarized, so even
this basic measurement requires a basic understanding of polarization.
Astronomers whose vision is so limited should read \S \ref{essentialone}
and \ref{hybridtwo}, and then turn to some other activity.

	Astronomers who are interested in astronomical magnetic fields
and esoteric scattering geometries need to go further and measure
polarization.  Magnetic fields are a very important force in
astrophysics, rivalling gravity and gas pressure in some regions such as
interstellar space. Some extragalactic edge-on disks obscure light from
the central black hole, but the highly polarized scattered light reveals
not only the radiation from the black hole but also the properties of
the scattering medium. 

	Synchrotron radiation is linearly polarized perpendicular to the
magnetic field with fractional polarization typically $\sim 70\%$;
pulsars are mainly linearly and partly circularly polarized;  Faraday
rotation, caused by the intervening magnetoionic gas, rotates the
position angle of  linear polarization; weak Zeeman splitting of
spectral lines produces two circularly polarized components, and strong
splitting also produces linear polarization; scattered spectral lines
and continuum radiation are linearly, and sometimes weakly circularly,
polarized.

	The basic reference for our discussion of the fundamentals is
the excellent book on astronomical polarization by Tinbergen (1996) and
references therein.  A more mathematical and fundamental reference is
Hamaker, Bregman, and Sault (1996), which the theoretically-inclined
reader will find of interest. Our discussion below will be heuristic in
nature, avoiding proofs and mathematical detail. We will make several
unproven statements and assertions; the explanations and justifications
can be found in the abovementioned references. Practical details of
calibration and application to real telescopes are in the series of
Arecibo Observatory Technical and Operations Memos by Heiles and his
collaborators and, also, in a forthcoming set of articles in the PASP;
all of these are listed in the references.
 
\section{The Jones vector for the electric field}

\subsection{ Polarization of an oscillating telephone cord}

	It's fun to take a long coiled telephone cord, tie one end to
a fixed point, and wiggle the other end to excite standing waves. These
waves are characterized by polarization, just as electromagnetic waves
are. 

\begin{figure} 
\plottwo{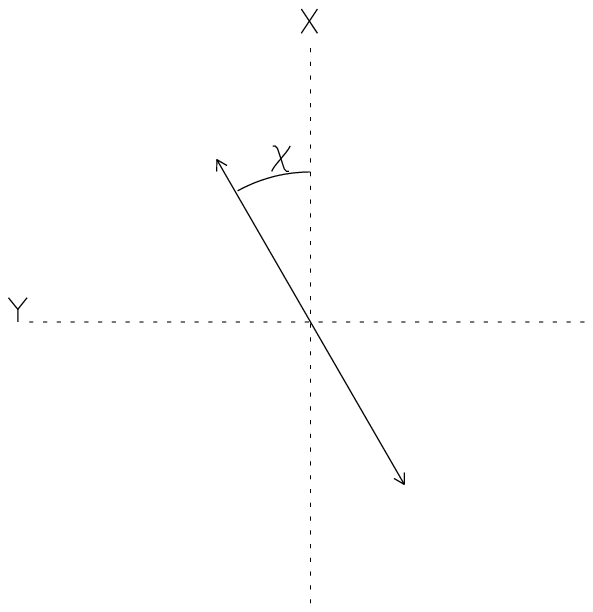}{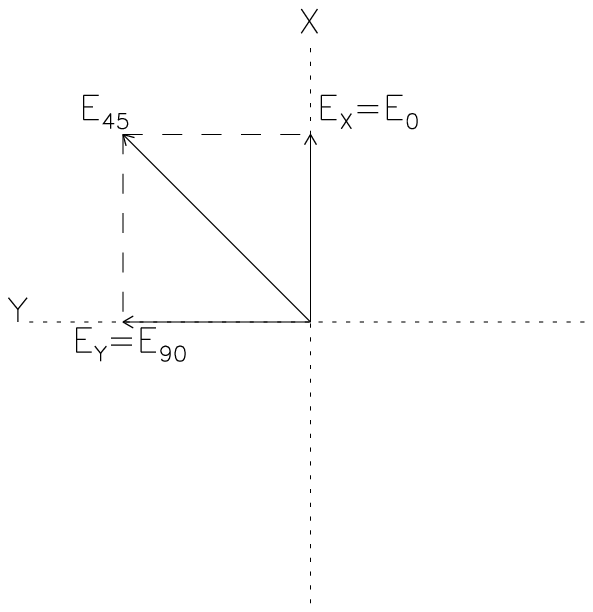}
\caption{{\it (a, left)}, linear polarization at position angle $\chi$;
{\it (b, right)}, $E_{45}$ is equivalent to the (non-vector) sum $(E_X +
E_Y)/\sqrt 2$. \label{vectordiagram} } 
\end{figure}

	If you wiggle back-and-forth vertically, it's vertically {\it
linearly} polarized, and ditto with horizontally. Let's call these
directions $X$ and $Y$, as in Figure \ref{vectordiagram}a. (Contrary to
usual, we denote the horizontal direction with $Y$!). More generally,
define the {\it position angle of polarization} $\chi$ to be measured
from the vertical in the counterclockwise direction; then when you
wiggle at angle $\chi$, the position angle of linear polarization is
also $\chi$. The peak amplitude in the $X$ direction is $A_{pX} = A_p
\cos \chi$ and in the $Y$ direction $A_{pY} = A_p \sin \chi$. So
generally, your wiggling produces amplitudes in both directions. Of
course, the stronger you wiggle, the larger the amplitude $A_p$. So your
wiggling in linear polarization mode can be specified by two quantities,
the amplitude and the position angle.

	$\chi$ is periodic in $\pi$, not $2 \pi$ like an ordinary angle.
This means that linear polarization doesn't have a {\it direction};
rather, it has an {\it orientation}. Often one represents its position
angle on the sky by short lines, like iron filings. These lines are
sometimes called vectors---incorrectly, because vectors do have a
direction.

	You can be fancier and wiggle with a circular motion, and this
can be either clockwise or counterclockwise. This is {\it circular
polarization}, and the two directions are called left-hand and
right-hand circular. Here, too, you have amplitudes in both directions.

	In what basic essence does the circular mode differ from the
linear mode at $\chi = 45\deg$? It's not that the $X$ and $Y$ directions
have different {\it amplitudes}. Rather, they have different {\it
phases}. Specifically, suppose you describe your vertical wiggle with 
$A_{px} = A_p \cos (\omega t)$; then you'd describe the horizontal
wiggle with  $A_{pY} = A_p \sin (\omega t)$. One's a cosine, the other a
sine. Now, there is only one difference between a cosine and a sine:
it's the phase angle. Remember the trig identity

\begin{equation}
\sin (\omega t) = \cos \left(\omega t - {\pi \over 2}\right)
\end{equation}

\noindent In other words, you can think of the $Y$ and $X$ amplitudes as
being identical in all respects except that they differ in phase by
$90^\circ$.  The phase angle can be positive or negative: positive
produces clockwise rotation, negative anticlockwise. More generally, the
$X$ and $Y$ two amplitudes can differ in phase by an arbitrary angle
$\phi$; when $\phi \neq 0^\circ$ or $90^\circ$, we have {\it elliptical}
polarization. Elliptical polarization can also be described by
different $X$ and $Y$ amplitudes.

\subsection{ Vectorial representation with trig notation for time}

	When you wiggle with linear polarization, you can describe $X$
and $Y$ amplitudes with identical trig functions, like $A_Y(t) = A_{pY}
\cos (\omega t)$ and $A_X(t) = A_{pX} \cos (\omega t)$. It's convenient
to ignore the time dependence and write the peak amplitudes using vector
notation:

\begin{eqnarray} 
\label{vectordef1}
{\bf A_p} = \left[ 
\begin{array}{c} 
    A_{pX} \\ A_{pY} 
\end{array} 
\; \right] \; .
\end{eqnarray} 

\noindent and in this case, with $A_{pX} = A_p \cos \chi$ and $A_{pY} =
A_p \sin \chi$, we write

\begin{eqnarray} 
{\bf A_p} = A_p \left[ 
\begin{array}{c} 
    \cos \chi \\ \sin \chi
\end{array} 
\; \right] \; .
\end{eqnarray} 

\noindent The meaning of this is perfectly clear without explicitly
writing the time dependence. However, suppose we are dealing with
circular polarization. Then, with ordinary trig notation, we need to
explicitly include the time dependence and write

\begin{eqnarray} \label{plusphi}
{\bf A(t)} = \left[ 
\begin{array}{c} 
    A_{pX}\cos (\omega t) \\ A_{pY} \cos (\omega t + \phi) 
\end{array} 
\; \right] \; .
\end{eqnarray} 

\noindent which is awkward: our description of the wiggling doesn't
really need the time dependence, which only specifies in a cumbersome
way that the motion is periodic with a certain frequency. What's really
important is the phase angle. To reflect this, often an equation like
the above is simplified by writing

\begin{eqnarray} \label{trigphase}
{\bf A_p} = \left[ 
\begin{array}{c} 
    A_{pX} \\ A_{pY} \angle \phi) 
\end{array} 
\; \right] \; .
\end{eqnarray} 

\noindent The meaning of this is clear, but mathematically one can't
manipulate the angle sign so it isn't useful in a mathematical sense.

\subsection{Complex exponential notation}

	Complex exponential notation is handy because it eliminates the
need to explicitly write the time dependence in terms of $(\omega t)$,
it allows one to specify phase angles, and the notation can be
mathematically manipulated. We recall that the complex plane has a real
and imaginary axis. The cosine and sine functions are the projections of
a complex exponential on the real and imaginary axis, respectively. We
have

\begin{equation}
e^{i \theta} = \cos \theta + i \sin \theta
\end{equation}

\noindent which leads to 

\begin{equation}
\cos \theta = Re \ [e^{i \theta}]
\end{equation}

\begin{equation}
\sin \theta = Im  \ [e^{i \theta}] = Re \ [-i e^{i \theta}]
\end{equation}

\noindent and, most importantly it allows the {\it addition} of a phase
angle to be replaced by the {\it multiplication} of its complex
exponential, 

\begin{equation}
\cos (\theta + \phi) = Re \ [e^{i (\theta + \phi)}] = 
Re \ [e^{i\theta} e^{i \phi} ]
\end{equation}

\begin{equation}
\sin (\theta + \phi) = Re \ [-ie^{i (\theta + \phi)}] = 
Re \ [-ie^{i\theta} e^{i \phi} ]
\end{equation}

	This last is important for our purposes because we can use
exponential notation to replace equation \ref{plusphi} with

\begin{eqnarray} \label{expphase}
{\bf A} = \left[ 
\begin{array}{c} 
    A_{pX} \\ A_{pY} e^{i \phi} 
\end{array} 
\; \right] \; ,
\end{eqnarray} 

\noindent which tells the essence, namely that $A_Y$ lags $A_X$ by phase
$\phi$.

	All aspects of the above discussion carry over to the electric
field in electromagnetic waves; we simply replace the amplitude $A$ with
the electric field $E$. Its vector representation is called the {\it
Jones vector}.

\section{ Polarized power and the Stokes vector}

	In astronomy we are almost never interested in the electric
field because we measure the {\it power}. Power is the time average of
the square of the electric field. Or, rather, the time average of the
product of the electric field with its complex conjugate; this takes
care of any difficulties with phases. 

\subsection{ How many parameters are required?}

	We made a heuristic argument above that the polarization is
specified by three parameters, namely the three in equation
\ref{expphase}.  However, there is one additional wrinkle. We were
describing a $100\%$ polarized wave in which the amplitude has a single
frequency and, correspondingly, a single polarization mode, which is
generally elliptical. Any monochromatic wave exists forever, so its
polarization never changes.

	However, there can also exist a superposition of sine waves of
different frequencies within some bandwidth. In fact, this is {\it
always} the case of natural radiation like sunlight, the 21-cm line, or
even astronomical masers. In nature, they are packed tightly together in
frequency with infinitely small separations. They produce an electric
field that varies randomly with time. These fields can all be polarized
in the same sense, just like a monochromatic wave. 

	But the polarizations can also be randomly distributed with
frequency. In this case we have unpolarized radiation. The Jones vector,
which treats a single sine wave and has only three parameters, is not
adequate to describe this case. 

	This extra possibility, that the electric field can have a
time-random unpolarized component, turns the three parameters into four:
the fourth tells the fraction of power that is nonpolarized. These four
parameters are combinations of polarized power called {\it Stokes
parameters}.  

	Stokes parameters are linear combinations of power measured in
{\it orthogonal polarizations}. We measure power in a particular
polarization by constructing an antenna that responds to that
polarization, meaning that the incoming electric field generates a
corresponding voltage in a cable. With a radio astronomy ``dish'', the
antenna is called a {\it feed}. Here we will think of a feed's antenna
as a probe in a waveguide that extracts linear polarization. More
generally, feeds can be made to sample linear, circular, or even
arbitrary elliptical polarization.

	We describe radiation in terms of electric fields having
particular polarizations; what we mean is that we have constructed an
antenna sensitive to that polarization and when we write the electric
field $E$ we really mean voltage in the cable that was induced by the
field. 

\subsection{ Linear polarization: Stokes Q and U}

	It's intuitively obvious that orthogonal {\it linear}
polarizations have $\chi$ differing by $90^\circ$: for example, vertical
($X$) and horizontal ($Y$) polarizations are orthogonal and have $\chi =
(0^\circ, 90^\circ)$, respectively.

	The power in the $(X,Y)$ directions is just $(E_X^2, E_Y^2)$,
respectively\footnote{To be more precise, one must realize that the
$E$'s have phases and are therefore complex, so the proper expressions
for power are $(\langle E_X \overline E_X \rangle, \langle E_Y \overline E_Y
\rangle)$, where the $\langle \rangle$ denotes time averages and the bar
denotes complex conjugate. We ignore these complications in this
introductory portion.}. We can linearly combine these two powers by
adding and subtracting them, and in the process we generate the first
two Stokes parameters $I$ and $Q$:

\begin{equation} \label{stokesi}
I = E_X^2 + E_Y^2 = E_{0^\circ}^2 + E_{90^\circ}^2
\end{equation}
\begin{equation} \label{stokesq}
Q = E_X^2 - E_Y^2 = E_{0^\circ}^2 - E_{90^\circ}^2
\end{equation}

\noindent Let us reflect on these quantities for a moment. 

\subsubsection{ The sum: Stokes I}

	The {\it sum} represents the {\it total power} in the incoming
radiation. It makes intuitive sense that, by sampling two orthogonal
polarizations, you pick up all of the incoming power.  It may not make
intuitive sense, but is nevertheless true, that it doesn't matter {\it
which} two orthogonal polarizations you measure and add together. The
orthogonal circulars, or two orthogonal linears at any pair of angles
$(\chi, \chi + 90^\circ)$, or even orthogonal ellipticals always sample
all of the power and their summed powers always gives the total
intensity $I$. 

	This is easy to see for the particular case of linear
polarization at $\chi = 45^\circ$. You can express $E_{0^\circ}$ in
terms of $(E_{45^\circ}, E_{-45^\circ})$ (see Figure
\ref{vectordiagram}b):

\begin{equation} \label{equalityone}
E_{0^\circ} = { E_{45^\circ} + E_{-45^\circ} \over \sqrt 2}
\end{equation}
\begin{equation} \label{equalitytwo}
E_{90^\circ} = { E_{45^\circ} - E_{-45^\circ} \over \sqrt 2}
\end{equation}

\noindent and when you take the sum of the squares, you find---naturally
enough---that  $E_{0^\circ}^2 + E_{90^\circ}^2 =  E_{45^\circ}^2 +
E_{-45^\circ}^2$.

\subsubsection{ The difference: Stokes Q}

	The difference tells about the polarization. Suppose the
incoming electric field is vertically polarized (the $X$ direction);
then $Q = I$. If it's horizontally polarized, then equation
\ref{stokesq} says $Q = -I$. If it's coming in at $\chi = 45^\circ$,
then $Q=0$. In fact, more generally, 

\begin{equation} \label{trigq}
{Q \over I} = p_{QU} \ \cos (2 \chi)
\end{equation}

\noindent where $p_{QU}$ is the total fractional linear polarization
(which we discuss below). So $Q$ is a very valuable diagnostic of the
linear polarization! But also it's {\it not complete}: for example, for
$\chi = 45^\circ$ we have $Q=0$ and, with this parameter alone, we would
not suspect that the signal is polarized.

\subsubsection{ Another difference: Stokes U}

	We need one more parameter to completely define the linear
polarization. That parameter is Stokes $U$, and is equal to

\begin{equation}
U =  E_{45^\circ}^2 - E_{-45^\circ}^2
\end{equation}

\noindent With a little reflection it becomes clear that

\begin{equation} \label{trigu}
{U \over I} = p_{QU} \ \sin (2 \chi)
\end{equation}

\noindent The combination $(Q,U)$ completely specifies the linear
polarization of the signal. The combination $(Q^2 + U^2)^{1/2}$ is the
total linearly polarized power and is independent of $\chi$. Generally,
even for a partially polarized signal, the fraction of linear
polarization and its position angle are given by

\begin{equation}
p_{QU} = \left[ \left({Q \over I}\right)^2 + \left({U \over I}\right) ^2  \right]^{1/2}
\end{equation}
\begin{equation}
\chi = 0.5 \tan^{-1} {U \over Q}
\end{equation}

\subsection{ Circular polarization: Stokes V}

	There are only two circular polarizations, which are called
right- and left-handed, or RCP and LCP, and they are orthogonal. So one
can derive two Stokes parameters from them: one is $I$, which is the
same as discussed above; you can include the $90^\circ$ phase difference
in equations such as \ref{equalityone} and \ref{equalitytwo} prove this
for yourself. 

	The difference is Stokes $V$. If you ever work with circular
polarization, you have to be careful about sign. Physicists use one
definition, radio astronomers use another (the IEEE definition,
reflecting our EE heritage), and optical astronomers use both, sometimes
without bothering to specify exactly which they are using. The IEEE
definition is

\begin{equation} V = E_{LCP}^2 - E_{RCP}^2 
\end{equation}

\noindent $LCP$ is generated by transmitting with a left-handed helix,
so from the transmitter the E vector appears to rotate anticlockwise. 
From the receiver, $LCP$ appears to be rotating clockwise.

	We define the fractional circular polarization just as we do for
linear polarization:

\begin{equation}
p_V = {V \over I} 
\end{equation}

\noindent $V$ can be positive or negative, and one can retain the sign
in the definition of $p_V$ if one wishes, as we've done here. 

\subsection{ The Stokes vector and total polarized power}

	We have four Stokes parameters, and it will be convenient to
write them in vector format, the 4-element {\it Stokes vector} 

\begin{eqnarray} 
\label{IQUV}
{\bf S} = \left[ 
\begin{array}{c} 
    I \\ Q \\ U \\ V \\
\end{array} 
\; \right] \; .
\end{eqnarray} 

\noindent The total fractional polarization is just

\begin{equation}
p = \left[ \left({Q \over I}\right)^2 + 
	\left({U \over I}\right) ^2 + 
        \left({V \over I}\right) ^2  \right]^{1/2}
\end{equation}

\noindent If both $p_{QU}$ and $p_V$ are nonzero, then the polarization
is {\it elliptical}, which is the general situation. Every elliptical
polarization has its orthogonal counterpart, and one can even build an
elliptically polarized feed. One normally prefers pure linear or
circular and tries to avoid the intermediate cases. However, Arecibo
uses turnstile junctions for some receivers, which have the advantage
that the polarization can be adjusted to pure circular with exquisite
accuracy---but the polarization becomes increasingly elliptical,
changing to linear and back again, with increasing departure from the
design frequency (see Heiles et al 2000b, 2001b)!

\subsection{If you don't remember anything else, remember THIS!}
\label{essentialone}

	Often you find yourself needing to combine polarizations. For
example, if you measure the polarization of some object several times,
you need to average the results. 

	There is only one {\it proper} way to combine polarizations, and
that is to use the Stokes parameters. The reason is simple: because of
conservation of energy, powers add and subtract. But it is definitely
wrong to average fractional polarizations or angles. Consider that
fractional polarizations are always positive, so they can never average
to zero. And angles are even worse! Consider averaging two measurements
having angles of $0\deg$ and $179\deg$---angles that differ by only
$1\deg$  because of the $\pi$ periodicity of position angle. The
straight average of the angles gives about $90\deg$---the orthogonal
polarization! 

	What you must do is convert the fractional polarization and
position angle to Stokes parameters, average the Stokes parameters, and
convert back to fractional polarization and position angle.

\section {Measuring Stokes parameters in radio astronomy}

	In contrast to optical astronomers, radio astronomers can
measure all Stokes parameters {\it simultaneously}. It may not be
obvious how they do this: we've described Stokes parameters as
differences between powers in various pairs of orthogonal polarizations,
each pair ``belonging'' to a particular Stokes parameter, and we can't
simultaneously place six feed probes at the same physical location to
simultaneously measure $(E_{0^\circ}, E_{90^\circ}, E_{45^\circ},
E_{-45^\circ}, E_{LCP}, E_{RCP})$ because all these antennas would
interact with other and make a total mess. Fortunately, we can generate
a Stokes parameter not only by subtraction of its own orthogonal
polarizations, but also by {\it multiplying} electric fields of two
orthogonal polarizations belonging to a {\it different} Stokes
parameter\footnote{We don't have to multiply; we can add and subtract,
as in Figure \ref{vectordiagram}b. But then we lose the advantage of
crosscorrelation discussed in \S \ref{correlated}}. 

\subsection{ Example: generating Stokes U from
$(E_{0^\circ},E_{90^\circ})$}

	This is easy to see for the case of deriving Stokes $U$ (which
is $E_{45^\circ}^2 -E_{-45^\circ}^2$) from it's non-belonging brethren
$(E_{0^\circ},E_{90^\circ})$. Referring again to Figure
\ref{vectordiagram}b, it is clear, graphically speaking, that the
product $(E_{0^\circ}E_{90^\circ})$ is related to Stokes $U$. As the E
field at $\chi = 45^\circ$, which belongs to Stokes $U$, oscillates, it
induces in-phase fields in the $0^\circ,90^\circ$ directions, each
smaller by a factor of $\sqrt 2$. The E-fields in these two directions
are therefore {\it correlated}. If you measure the time average product
$\langle E_{0^\circ}E_{90^\circ}\rangle$, it's identical to 
$\langle E_{45^\circ}^2\rangle \over 2$. 

	In particular, if you begin with equations \ref{equalityone} and
\ref{equalitytwo} you can easily show that

\begin{equation} 
U \equiv E_{45^\circ}^2 - E_{-45^\circ}^2 = 2E_{0^\circ}E_{90^\circ}
\end{equation}

\noindent If you throw in a phase factor of $90^\circ$ in the above
equation, you'll recover Stokes $V$---this makes sense because the only
difference between linear and circular polarization is, in fact, the
phase factor. 

\subsection{ A general expression for Stokes parameters}

	One can write Stokes parameters in terms of electric fields of
any two orthogonal polarizations. Here we provide the version in which
one measures $(E_X, E_Y)$ with linearly polarized antennas at $\chi =
0^\circ$. For this case, 

\begin{equation} \label{Sdefinitionone}
I = E_X \overline{ E_X} + E_Y \overline{ E_Y}
\end{equation}
\begin{equation}
Q = E_X \overline{ E_X} - E_Y \overline{ E_Y}
\end{equation}
\begin{equation}
U = E_X \overline{ E_Y} + \overline{ E_X} E_Y 
\end{equation}
\begin{equation}
\label{Sdefinitionfour}
iV = E_X \overline{ E_Y} - \overline{ E_X} E_Y \; .
\end{equation}

\noindent The overbar indicates the complex conjugate. These products
are time averages; we have omitted the indicative $\langle \rangle$
brackets to avoid clutter. And remember, these equations apply to the
voltages induced into the antenna as well as to the original electric
fields, because they are proportional; below, we are always referring to
voltages even though we will write $E$.

	These equations make it clear how to measure all four Stokes
parameters simultaneously. Namely, begin with orthogonal polarizations;
any pair will do, but our equations are written for orthogonal linears.
Then digitize the resulting voltages and perform the above products in a
computer. The aspects of digitizing and computing are a story all in
themselves, but we leave that for another time.

\subsection{ The need for calibration}

	The above equations \ref{Sdefinitionone}-\ref{Sdefinitionfour}
are simple in theory but not so simple in practice because the
radioastronomical receiving system produces unwanted modifications in
the astronomical polarization. Feeds are almost never perfect, so their
polarizations are only approximately linear or circular; and generally
speaking, no feed has two outputs that are perfectly orthogonal. 

\begin{figure} 
\plottwo{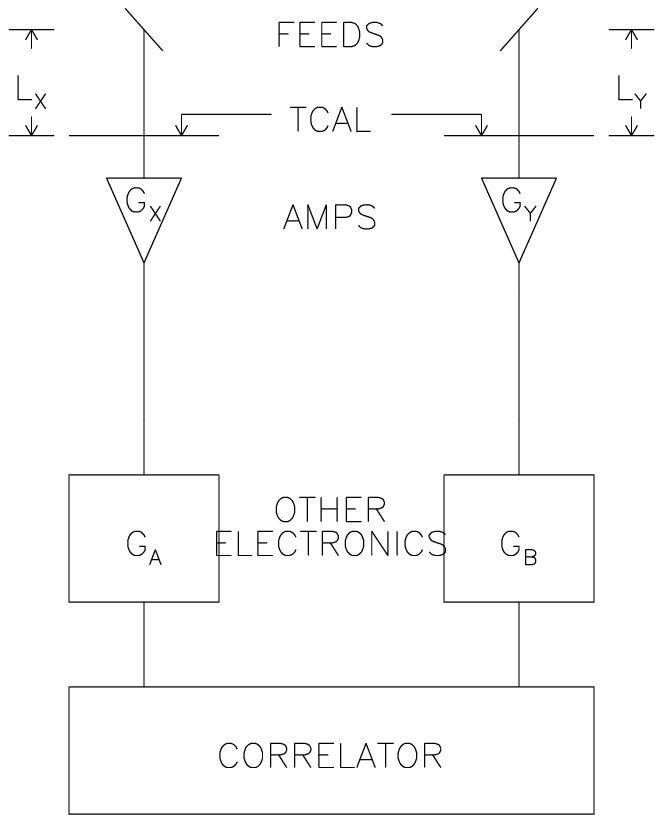}{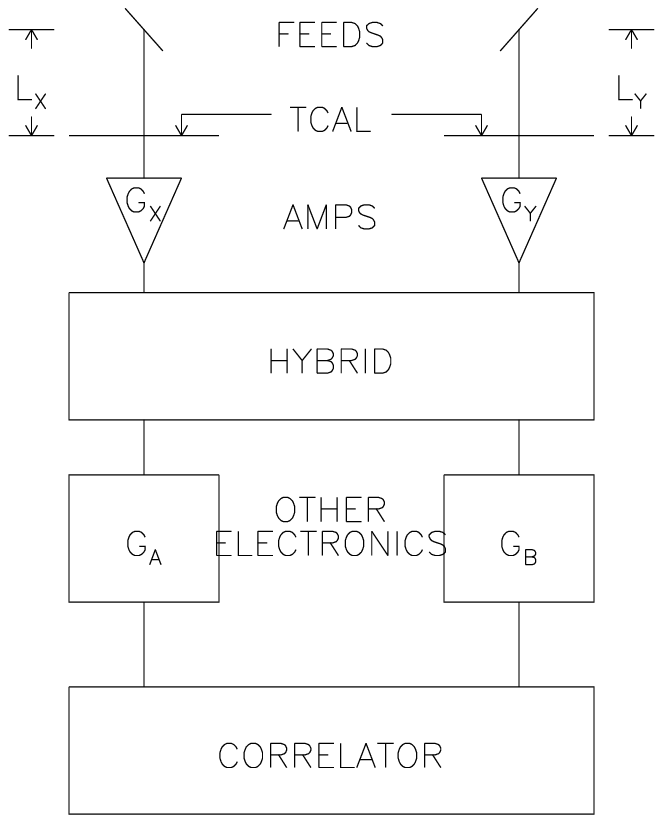}
\caption{Two versions of a radio astronomical receiver: {\it (a,left)},
{\bf good}; {\it (b, right)},
{\bf bad} (see \S \ref{hybridone} and \ref{hybridtwo}).
\label{rcvr}}
\end{figure}

	Most important in practice is the electronics system, which
introduces its own relative gain and phase differences between the two
linearly polarized channels.  Figure \ref{rcvr}a shows the important
elements of the system for this discussion. Two orthogonal feed probes
in a waveguide convert the incoming electric field to voltages. These
travel through cables having lengths $(L_X,L_Y)$ to a directional
coupler, where the correlated noise source is injected (through cables
of different length---a fact we ignore for the sake of simplicity).
$(L_X,L_Y)$ cannot be exactly identical, and the difference produces a
phase shift between the two polarization channels. The two polarization
channels are amplified with gains that are inherently complex, meaning
that each introduces its own phase shift. The signals are multiplied in
a mixer by a local oscillator, again injected through cables whose
lengths cannot be precisely equal, leading to an additional phase offset
between the channels. The resulting i.f.\ signals are sent down to the
digital correlator through cables from the feed; these also have
different lengths and losses. 

	The total gains in the (left, right) channels are
$(G_XG_A,G_YG_B)$. If the {\it magnitudes} of these gains differ, then
the difference between the two channels is nonzero for an unpolarized
source, making the source appear to be {\it linearly} polarized.  If the
{\it phases} differ, then a linearly polarized source appears to be {\it
circularly} polarized.  These electronics gains and phases must be
calibrated relatively frequently because they can change with time.  
This calibration is most effectively done by injection of a correlated
noise source into the two feed outputs. 

\section{ Calibration: Jones and Mueller matrices}

	The {\it modification} of the derived polarization by the 
system components is most generally described by matrices.  The {\it
Jones matrix} describes the modification of the Jones vector, and the
{\it Mueller matrix} describes the modification of the Stokes vector.  

	Return to our wiggling telephone cord. Suppose we wiggle one end
with linear polarization in the vertical direction, making $Q=I$ and
$(U,V)=0$. Now a Martian spy starts to move the other end around in a
small circle at the same frequency. This would change the original pure
vertical linear polarization into elliptical polarization, taking some
power from $Q$ and putting it into $U$ and $V$.

	Feeds and electronic devices are like the Martian, modifying the
electric field's polarization. Generally, a device can couple an
arbitrary fraction of the $Y$ voltage into $X$ with an arbitrary phase;
and also {\it vice-versa}.  The easy way to write these mutual
perturbations is with a matrix transfer function that relates the output
voltages to the input voltages. This matrix operates on the Jones
vector, so it is naturally called the {\it Jones matrix}. With two
orthogonal polarizations, there are two voltages; thus the Jones matrix
is $2 \times 2$. 

	If the Jones matrix is unitary, then it produces no modification
in the original polarization. Unitary matrices aren't very interesting in
astronomy, because we need lots of amplification for the tiny voltages
induced at the feed! But we can imagine a system in which the Jones
matrix is diagonal; that would mean there is no coupling between the $X$
and $Y$ channels. Large diagonal elements would increase the voltages so
that we can measure them, and if the two diagonal elements were equal,
it would also keep the polarization---and thus the Stokes
parameters---unchanged.

	More generally, the Jones-matrix-modified voltages change the
polarization state, and thus change the Stokes parameters.  So every
Jones matrix has its Stokes-parameter counterpart, which is called the
{\it Mueller matrix}. The Mueller matrix relates the output Stokes
vector of equation \ref{IQUV} to the input Stokes vector:

\begin{equation}
\label{transfereqn}
{\bf S_{out}} = {\bf M \cdot S_{in}} \; .
\end{equation}

\noindent There are four Stokes parameters, so the Mueller matrix is $4
\times 4$; in general, all elements may be nonzero, but they are not all
independent. In the usual way, we write

\begin{eqnarray}
\label{Mueller}
{\bf M} = \left[
\begin{array}{cccc}
   m_{II}   & m_{IQ}   & m_{IU}   &  m_{IV} \\
   m_{QI}   & m_{QQ}   & m_{QU}   &  m_{QV} \\
   m_{UI}   & m_{UQ}   & m_{UU}   &  m_{UV} \\
   m_{VI}   & m_{VQ}   & m_{VU}   &  m_{VV} \\
\end{array}
\; \right] \; .
\end{eqnarray}  

\noindent Each matrix parameter is the coupling of the two Stokes
parameters indicated by its subscripts. 

	Each system component has its own Jones and Mueller matrices,
and the total system matrices are the products of the individual
component's matrices. The Mueller matrix for the feed has complicated
off-diagonal elements. Fortunately, feeds are usually well-designed
and these off-diagonal elements are small, meaning the instrumental
polarization is small. 

	Here, we will restrict our detailed discussion to the two most
important specific components, the electronics chain and the coupling of
the telescope to the sky; fortunately they are also the simplest. For a
complete treatment that includes the feed, see Heiles et al (2000a,
2000b). 

\subsection{ The Jones and Mueller matrices for the electronics chain}

	The two polarization channels go through different amplifier
chains as in figure \ref{rcvr}a, and we can safely assume that
there is no coupling between the channels in the electronics system;
this, in turn, means that the nondiagonal elements of the Jones matrix
are zero. Suppose the two channels have {\it voltage} gain $(g_A, g_B)$,
{\it power} gain $(G_A, G_B) = (g_A^2, g_B^2)$, and phase delays
$(\psi_A, \psi_B)$. Clearly, the Jones matrix is

\begin{eqnarray}
\label{jonesmatrix2}
\left[
\begin{array}{c}
   E_{A,out}  \\
   E_{B,out}  \\
\end{array}
\; \right] = \left[
\begin{array}{cc}
   g_A e^{i \psi_A}      &            0                   \\
           0             &      g_B e^{i \psi_B}          \\
\end{array}
\; \right] \left[
\begin{array}{c}
   E_{A,in}  \\
   E_{B,in}  \\
\end{array}
\; \right] \; 
\end{eqnarray}

	You calculate the Mueller matrix $\bf M_A$ from the Jones matrix
by the laborious procedure of applying equations
\ref{Sdefinitionone}-\ref{Sdefinitionfour}. The result is surprisingly
complicated. However, for a well-designed and calibrated system, we can
assume that the two gains $(g_A,g_B)$ are nearly identical, and for
simplicity assume that their average is unity. That is, we can assume
that

\begin{equation}
\Delta G \equiv G_A - G_B \ll 1
\end{equation}

\noindent We then carry the algebra only to first order in $g_Ag_B$,
meaning we take $g_Ag_B=1$.  With this first-order approximation, the
electronics Mueller matrix becomes

\begin{eqnarray} \label{mmtrx}
{\bf M_A} = \left[ 
\begin{array}{cccc} 
  1                &  {\Delta G \over 2}   &     0    &      0          \\
{\Delta G \over 2}  &          1           &     0    &      0          \\
  0                &       0              & \cos \psi & -\sin \psi  \\
  0                &       0              & \sin \psi & \cos \psi  \\
\end{array} 
\; \right] \; .
\end{eqnarray} 

\noindent Here we have set $\psi \equiv \psi_A - \psi_B$: the difference
is all that matters because it's always the {\it relative} phase between
the two channels that counts. The matrix consists of two submatrices,
the upper left and the lower right. Let's reflect on these submatrices.

	The upper left submatrix represents coupling between Stokes $I$
and $Q$. The relative gain error directly affects these two parameters
because they are the sum and difference of the $X$ and $Y$ powers. In
contrast, these powers are completely unaffected by the phase
difference, so $\psi$ doesn't appear in this submatrix. The two diagonal
elements are unity because we've assumed $\Delta G \ll 1$. Consider the
specific example of an unpolarized source: a gain difference makes it
look polarized, with $Q$ nonzero, and it also affects Stokes $I$.

	The lower right submatrix represents coupling between Stokes $U$
and $V$. The relative phase directly affects these two parameters
because they are the correlated products $E_XE_Y$, and the departure of
the correlation's phase angle from zero directly reflects the degree of
circular polarization, Stokes $V$. 

\subsection{A very important property of correlated voltages}
\label{correlated}

	Embodied in equation \ref{mmtrx} is a highly important
principle: {\it gain errors do not affect correlation products when
there is no polarization}. This is important because amplifier gains
fluctuate with time and their calibration is subject to measurement
error; in contrast, amplifier phase delays and cable lengths tend to
change only slowly with time.

	Consider the case of a nonpolarized source. Equation \ref{mmtrx}
shows that the error in the measured Stokes $Q$ is directly proportional
to  $\Delta G$. However, Stokes $U$ and $V$ are completely independent
of $\Delta G$, so a nonpolarized source {\it cannot} produce fake
nonzero $(U,V)$. This is because if $E_XE_Y = 0$, as it is for an
unpolarized source, then the product is zero even with a gain error. 

	In practice, this makes the measurement of small polarizations
more accurate when using correlated products. This fundamental fact
appears again and again in precision radioastronomical measurements of
small quantities, and is also the basis for interferometry. The
corollary is the somewhat nonintuitive fact: {\it If you want to measure
small circular polarizations accurately, then use linearly polarized
feeds; if you want to measure small linear polarizations accurately,
then use circularly polarized feeds!}

\subsection{Carrying correlation too far: using a hybrid} \label{hybridone}

	Suppose you want to measure weak linear polarization. As we
discussed above, the best technique is to crosscorrelate orthogonal
circulars. But suppose your telescope only has a linearly polarized
feed!

	You may be tempted to modify the system block diagram to include
a $90\deg$ hybrid, as shown in Figure \ref{rcvr}b. The hybrid
inserts a $90\deg$ phase shift into one channel and then adds them.
This turns the dual linear system [polarizations $(X,Y)$] into a dual
circular one [polarizations $(A,B$)]. You can then use the
crosscorrelation technique to generate the dual linears. 

	However, this system does {\it not} provide the abovementioned
advantage of crosscorrelation in measuring small signals. The reason is
simple: the $(X,Y)$ signals are combined {\it after} the first amplifier
and have been multiplied by the gains $(G_X,G_Y)$ with their
corresponding uncertainty and time variability. Thus, the combined
$(A,B)$ signals are pure circular polarization only to the extent that
$G_X = G_Y$---and, of course, this includes their complex portions, the
phases, as well. And we are ignoring the inevitable imperfections
in the hybrid. After the hybrid, the complex channel gains $(G_A,G_B)$
operate on the signal, as usual. 

	We now have {\it four} combinations of gain to worry about:
$( G_XG_A, G_XG_B$, $G_YG_A, G_YG_B)$. In contrast, the straightforward
system without the hybrid has only two combinations: $(G_XG_A,G_YG_B)$.
This makes the calibration process for the hybrid system more
complicated, requiring turning on the correlated cal not only when it is
connected to both channels simultaneously but also when it is connected
to each one individually, one at a time. The details are discussed by
Heiles and Fisher (1999). 

	The hybrid {\it completely} removes the cross correlation
advantage: with the hybrid, there is {\it no} Stokes parameter that is
unaffected by $\Delta G$. 

\subsection{Stokes I and the hybrid} \label{hybridtwo}

	The hybrid even creates problems measuring Stokes $I$ because of
the more complicated calibration procedure described above.
Unfortunately, however, astronomers rather traditionally prefer to
measure Stokes $I$ using dual circular polarization instead of dual
linear. The reason is partly scientific, partly historical.
Historically, the first receiving systems had only a single polarization
channel: it was hard and expensive enough to make a single low-noise
receiver, let alone two. Classical radio sources, such as quasars, often
exhibit significant linear polarization, but very little circular. Thus,
to obtain reliable, repeatable flux density measurements---particularly
at low frequency, where ionospheric Faraday rotation is important---the
single polarization of choice was circular. Similarly, in historic
single-polarization VLBI with its differing ionospheric Faraday
rotations at the different stations, circular polarization was
preferred. And finally, pulsars are more highly linearly than circularly
polarized. When faced with a single polarization system and sources that
are linearly polarized, the polarization of choice is circular because
one needs only to multiply the measured flux by two to get  Stokes $I$.

	This traditional emphasis on circular polarization persists in
dual-polarized receivers. Many astronomers who want to measure nothing
more than Stokes $I$, when faced with a dual-linear system, insert a
hybrid to convert the system to dual-circular. But they don't carry
through with the extra steps of calibration required. 

	To bring the point home that using a hybrid is inappropriate,
consider the extreme case when the $Y$-polarization amplifier is turned
off. The astronomer who uses a hybrid points the telescope to the source
of interest and sees both channels $(A,B)$ respond. Then
he\footnote{Sexism here is intentional: female astronomers are
presumably smart enough to avoid such foolishness.} turns on the cal and
sees both channels respond. He has {\it no idea} that one channel is
dead. He might wonder why the levels are 3 db lower than usual, but
astronomers usually don't pay attention to power levels, ascribing them
to the domain of the receiver engineer. 

	In historical times, radio astronomers often did use a hybrid to
generate the circular polarization(s) from a dual linear feed, but placed
the hybrid {\it before} the first amplifier. This is far better, because
then one is reduced to the simpler situation of having only two sets of
gains to determine, $(G_XG_A, G_YG_B)$. The problem with this approach
is that hybrids have some loss, and therefore introduce noise. In those
historical times receiver temperatures were high enough that this extra
noise could be tolerated. Today's receiver temperatures are too low for
this approach unless the hybrid is cooled. 

	The moral: {\it don't ever use a post-amp hybrid unless you {\bf
really need} to change the polarization for a special, specific purpose!}
To measure Stokes $I$, use the native feed polarization; calibrate and
measure the two polarization channels independently, and add the
results.

\subsection{ The Mueller matrix for the sky}

	A linearly polarized astronomical source has Stokes
$(Q_{src},U_{src})$ parameters defined with respect to the north
celestial pole (NCP). It's these quantities that we want to measure. 

	The first device encountered by the incoming radiation is the
telescope (${\bf M_{SKY}}$). These days, all major telescopes are alt-az
mounted,. This means that the feed mechanically rotates with respect to
the sky as the dish tracks the source. The angle of rotation  is called
the {\it parallactic angle}  $PA_{az}$. It is defined to be zero at
azimuth $0\deg$ and increase towards the east; for a source near zenith,
which is always the case at Arecibo, $PA_{az} \sim az$, where $az$ is
the azimuth angle of the source.  The Stokes parameters seen by the feed
are $(Q_{sky}, U_{sky})$; the conversion between ${\bf S_{src}}$ and
${\bf S_{sky}}$ is given by  the Mueller matrix\footnote{One can also
write the corresponding Jones matrix, should one so desire.}

\begin{eqnarray} 
\label{skymatrix}
{\bf M_{SKY}} = \left[ 
\begin{array}{cccc} 
 1 &     0     &     0    & 0 \\
 0 &  \cos 2PA_{az} & \sin 2PA_{az} & 0 \\
 0 & -\sin 2PA_{az} & \cos 2PA_{az} & 0 \\
 0 &     0     &    0     & 1 \\
\end{array} 
\; \right] \; .
\end{eqnarray} 

\noindent The central $2 \times 2$ submatrix is, of course, nothing but
a rotation matrix. $\bf M_{SKY}$ doesn't change $I$ or $V$, which makes
sense.

	For an equatorially-mounted telescope, the feed {\it doesn't}
rotate on the sky as the source is tracked. This fact might still be of
interest to optical astronomers, but with the demise of the last of the
great equatorial telescopes---the NRAO 140-footer---this fact recedes
into the fog of history for us radio astronomers.

\subsection{The total system Mueller matrix}

\label{mmsystem}

	Heiles et al (2000a, 2001a) define seven parameters that specify
the complete system Mueller matrix. Two of these refer to the cal gain
and phase with respect to the sky (equation \ref{mmtrx} above), four
refer to the ellipticity and nonorthogonality of the feed, and the
seventh is a rotation angle. In contrast, the $4 \times 4$ Mueller
matrix contains sixteen elements; not all of the elements are
independent. For illustrative purposes, we write the full system matrix,
without the sky correction, in terms of the first six parameters:

\small{
\begin{eqnarray*}
\left[
\begin{array}{cccc}
                    1        & (-2 \epsilon \sin\phi \sin2\alpha +
{\Delta G \over 2} \cos 2\alpha) &
  2 \epsilon \cos\phi &  (2 \epsilon \sin\phi \cos2\alpha + {\Delta G
\over 2} \sin 2\alpha) \\
{\Delta G \over 2}           &   \cos 2\alpha  &  0    &  \sin 2\alpha
\\
2\epsilon \cos(\phi + \psi)  & \sin 2\alpha \sin\psi & \cos\psi  &
-\cos2\alpha \sin\psi  \\
2\epsilon \sin(\phi + \psi)  & -\sin 2\alpha \cos\psi &  \sin \psi  &
\cos2\alpha \cos\psi
\end{array}
\; \right] 
\end{eqnarray*}
}

        {\boldmath $\Delta G$} is the error in relative intensity
calibration of the two polarization channels. It results from an error
in the relative cal values $(T_{calA}, T_{calB})$.

        {\boldmath $\psi$} is the phase difference between the cal and
the incoming radiation from the sky (equivalent in spirit to $L_X -
L_Y$ on our block diagram..

        {\boldmath $\alpha$} is a measure of the voltage ratio of the
polarization ellipse produced when the feed observes pure linear
polarization.

        {\boldmath $\chi$} is the relative phase of the two voltages
specified by $\alpha$.

        {\boldmath $\epsilon$} is a measure of imperfection of the feed
in producing nonorthogonal polarizations (false correlations) in the two
correlated outputs.

        {\boldmath $\phi$} is the phase angle at which the voltage
coupling $\epsilon$ occurs. It works with $\epsilon$ to couple $I$ with
$(Q,U,V)$.          

        {\boldmath $\theta_{astron}$} is the angle by which the derived
position angles must be rotated to conform with the conventional
astronomical definition.

\section{Calibrating and using the matrix parameters}

\subsection{ The role of the correlated cal}

	In practice, the amplifier gains and phases are calibrated with
a correlated noise source (the ``cal'').  Thus, our amplifier gains
$(G_A, G_B)$ in equation \ref{mmtrx} have nothing to do with the actual
amplifier gains.  Rather, they represent the gains as calibrated by
specified cal intensities, one for each channel.  If the {\it sum} of
the specified cal intensities is perfectly correct, then the absolute
intensity calibration of the instrument is correct for an unpolarized
source (i.e.\ Stokes $I$ is correctly measured in absolute units). 
Above, we have assumed  $G_A +G_B = 2$, which means that we are dealing
with fractional polarizations and neglecting the absolute calibration of
intensity.  

	The difference between the amplifier {\it phases} is also
referred to the cal.  Thus the phase difference $\psi = \psi_A - \psi_B$
represents the phase difference that exists between a linearly polarized
astronomical source and the cal and has nothing to do with the amplifier
chains. 

	We assume the cal to be constant. Thus, the measured Stokes
vector is referred to the cal. This means that all artifacts of the
electronics chain, which change with time, are removed by referring the
measured Stokes vector to the cal, which is constant in time. 

	It remains to relate the cal to the sky. This must be done by
astronomical observations that determine the Mueller matrix of the
calibrated system. In other words, the calibrated system's Mueller
matrix multiplies the incoming Stokes vector from the sky and produces
the measured Stokes vector. 

\subsection{ Determining the Mueller matrix astronomically}

	Astronomical radio sources exhibit signficant linear
polarization but usually negligible circular polarization. We determine
the matrix astronomically by tracking a linearly polarized radio source
over a wide range of parallactic angle $PA$. As $PA$ changes, Stokes $Q$
and $U$ from the source are modified by ${\bf M_{SKY}}$ in equation
\ref{skymatrix}. In contrast, any $PA$ dependence of the measured Stokes
$V$ must reveal nonzero matrix elements coupling Stokes $(Q,U)$ into
$V$, namely $(m_{QV},m_{UV})$ and their two counterparts. 

\begin{figure} 
\plotone{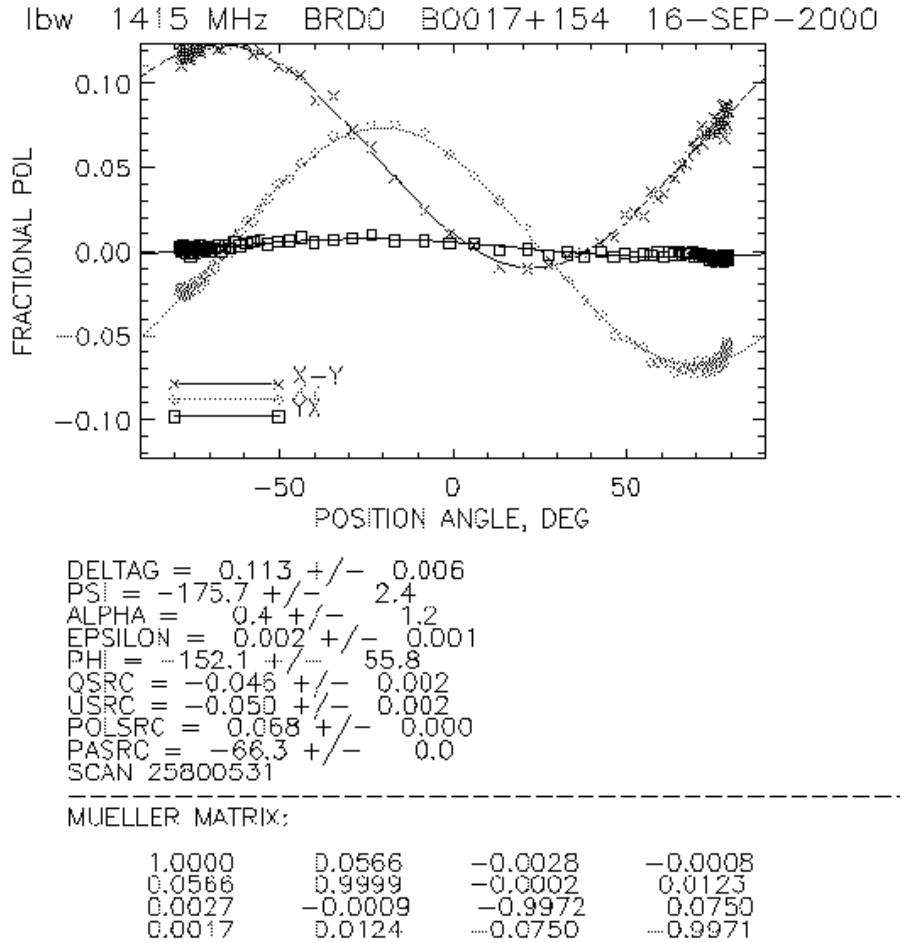}

\caption{A calibration observation of the source $0017+154$ for
Arecibo's dual-linear LBW feed. At the top, the measured Stokes
parameters, uncorrected for the system's and sky Mueller matrices,
versus $PA$. In the middle and bottom we have the derived values for the
instrumental parameters for the total Mueller matrix, and also the
Mueller matrix itself (\S {mmsystem}). \label{lbwfig}}

\end{figure}

	Figure \ref{lbwfig} shows a calibration observation vfor
Arecibo's dual-linear LBW feed. The crosses show the $PA$ dependence of
$X-Y$---the difference between the two linears, which is the measured
Stokes $Q$; the diamonds show $XY$, the measured Stokes $U$. They follow
sine and cosine waves with comparable amplitudes, as they should
(equations \ref{trigq}, \ref{trigu}). The smallness of the departure
from these conditions is expressed by the tiny values for $(m_{QU},
m_{UQ})$. The amplitude of the sine/cosine curves gives the linear
polarization of the source, $\sim 7\%$.

	However, the sine wave for the crosses is displaced above zero
by about 0.06. This reflects coupling of Stokes $I$ into $Q$, $m_{QI}$;
this is the effect of nonzero $\Delta G$, an error in the relative cal
values. In contrast, $m_{UI}$ is very small, consistent with its
derivation from crosscorrelation; the fact that $m_{UI} \neq 0$ reflects
cross coupling in the feed, which is described by the parameter
$\epsilon$ above. 

	Finally, Stokes $V$ exhibits a small $PA$ dependence, which
indicates nonzero values for $(m_{UV},m_{VU})$; this results from an
error in the relative phase of the cal with respect to the sky, $\Delta
\psi$. The small departure of Stokes $V$ from zero could result either
from nonzero $m_{VI}$ or from nonzero circular polarization of the
source; one needs to observe several sources to be sure.

\subsection{ Using the matrix to correct data}

	Once the system's matrix parameters is determined it is a simple
matter to correct the data: one simply multiplies the measured Stokes
vector by the inverse of the system Mueller matrix. This must, of
course, included the sky rotation portion. Matrices are noncommutatve
and you have to be careful about constructing the system Mueller matrix;
details are in Heiles et al (2000a, 2001a).

\subsection{ Jones and Mueller matrices for antenna arrays}

	When using an array of antennas, such as the VLA, the output of
each interferometer pair is equivalent to the output of the single dish
described above. The pair's measured Stokes parameters can be corrected
by a Mueller matrix, whose elements can be derived by tracking a
polarized source, in a similar manner to that described above. Each
antenna pair is characterized by a different Mueller matrix. Correcting
the output of each pair with its Mueller matrix is a {\it
baseline-based} scheme.

	However, for large arrays it is much more efficient to use an
{\it antenna-based} calibration scheme. With this, you characterize each
antenna by its Jones matrix; you obtain the Mueller matrix for any
antenna pair from  the two Jones matrices. The Jones matrices for the
individual antennas can be derived from the $PA$ dependence of the
Stokes vectors for all the baselines using a least square technique. 

\section{ Polarized beam structure}

	We've left unspecified the implicit fact that we've been
describing the Mueller matrix corrections on the axis of the main beam,
as we'd measure for a pulsar or a small radio source. You may be
surprised to learn that the telescope beam contains unavoidable,
intrinsic polarization structure. The type of structure depends on the
Stokes parameter. 

	The fundamental cause of the polarized structure is the curved
reflector surface, which slightly changes the direction of an incident
linearly polarized electric vector upon its reflection. On the main beam
axis these distortions cancel, but off-axis they don't. The distortions
add in fundamentally different ways for linear and circular polarization
because, when a source is off-axis, the path lengths to the source from
different portions of the reflecting surface are not all equal.  The
distortions increase with curvature, and hence become more serious with
decreasing focal ratio; beam squint varies $\left({f \over
D}\right)^{-1}$ (Troland and Heiles 1982). Radio telescopes have small
$f \over D$, so the effects become very significant. 

	We don't have the space to delve into the somewhat esoteric
details of these distortions; see Tinbergen (1996, \S 5.5.5) and quoted
references. These descriptions are usually given for prime-focus fed
paraboloids; with their different geometries, Arecibo and the GBT differ
in detail but not in fundamental principle. 

\subsection{Main beam linear polarization}

	With linear polarization, there are two sources of
distortion. One relates to the feed: for a feed probe sampling the $X$
direction, the waveguide nature of the feed tends to make the feed's
illumination pattern broader in $X$ than in $Y$. After reflection from
the dish surface, the telescope HPBW is broader in $Y$ than $X$. The
second is the abovementioned dish curvature, which also produces a
similar distortion. We call these differing beamwidths {\it beam
squash}.

	Both effects produce the same result, namely different
beamwidths in orthogonal polarizations. When these two polarizations are
subtracted to produce the Stokes $(Q,U)$ parameters, one obtains a
four-lobed ``cloverleaf''structure in the $(Q,U)$ beam responses.
Figure \ref{mkfig1iquv} shows an example, taken at
Arecibo for a source near the telescope's maximum zenith angle $20\deg$
where some additional distortions are introduced and the sidelobe is
somewhat accentuated. The main beam exhibits not only the expected
squash, but also squint and higher-order distortions (Heiles 1999;
Heiles et al 2000b, 2001b).

\begin{figure} 
\plottwo{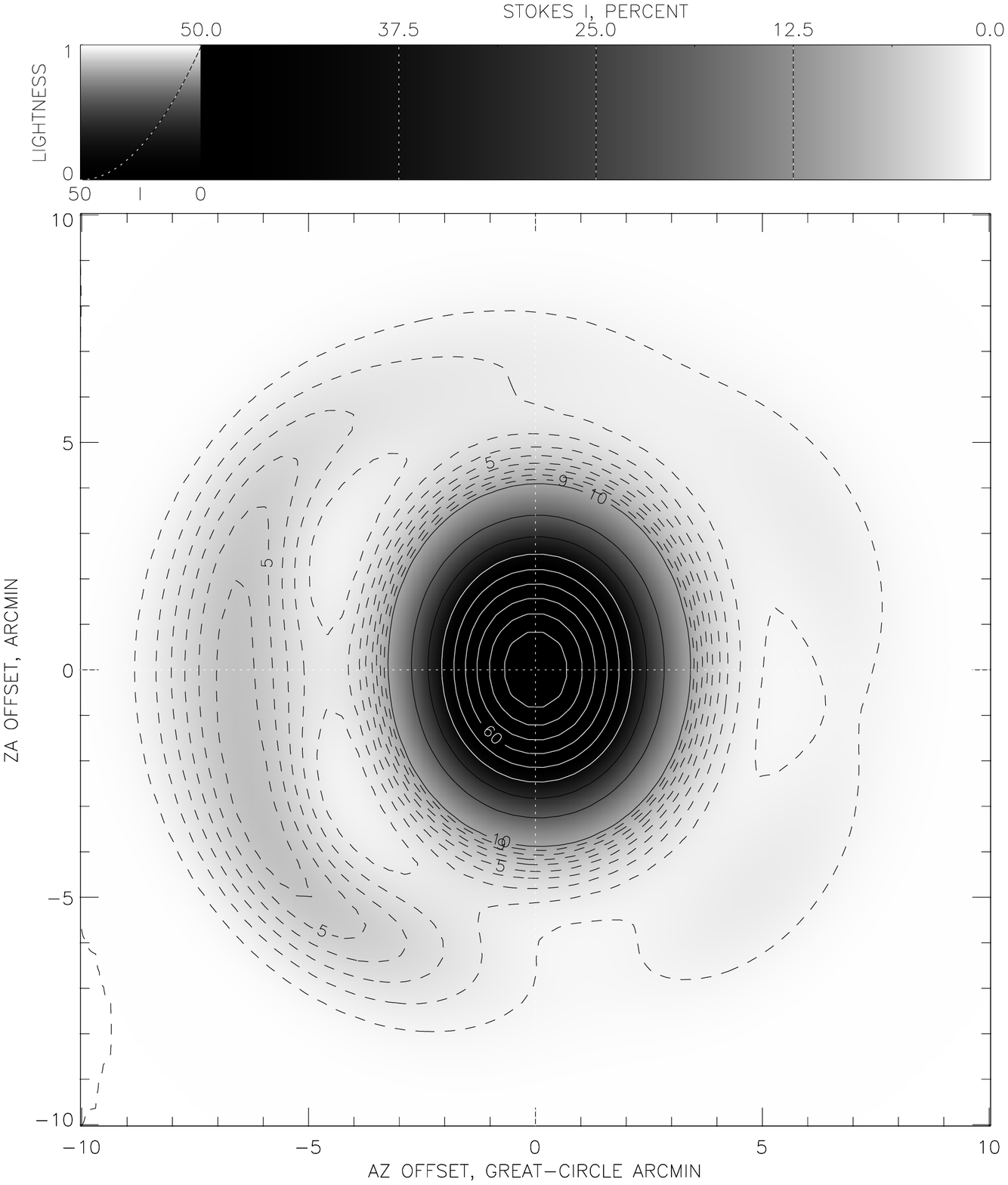}{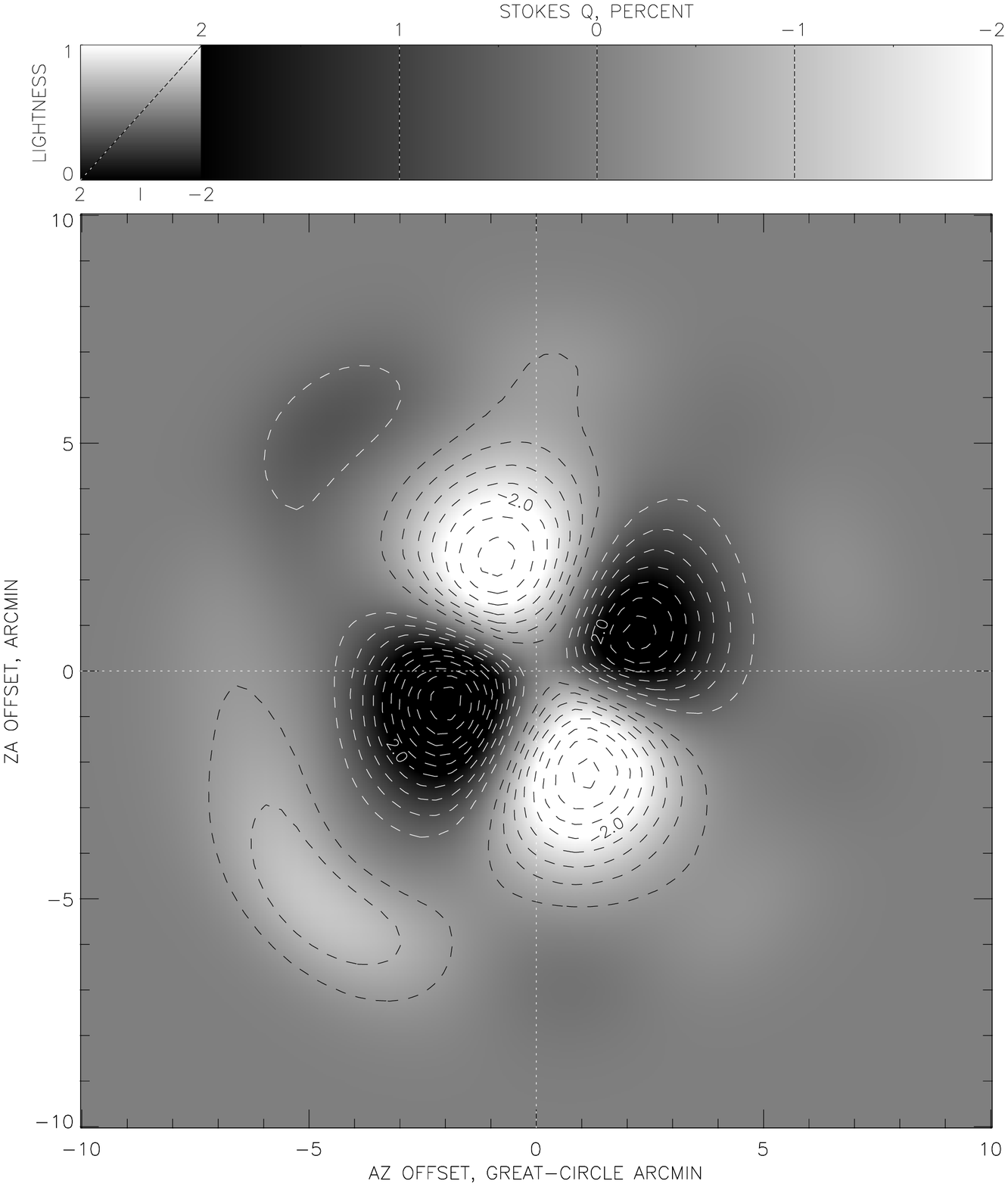}
\end{figure}
\begin{figure}
\plottwo{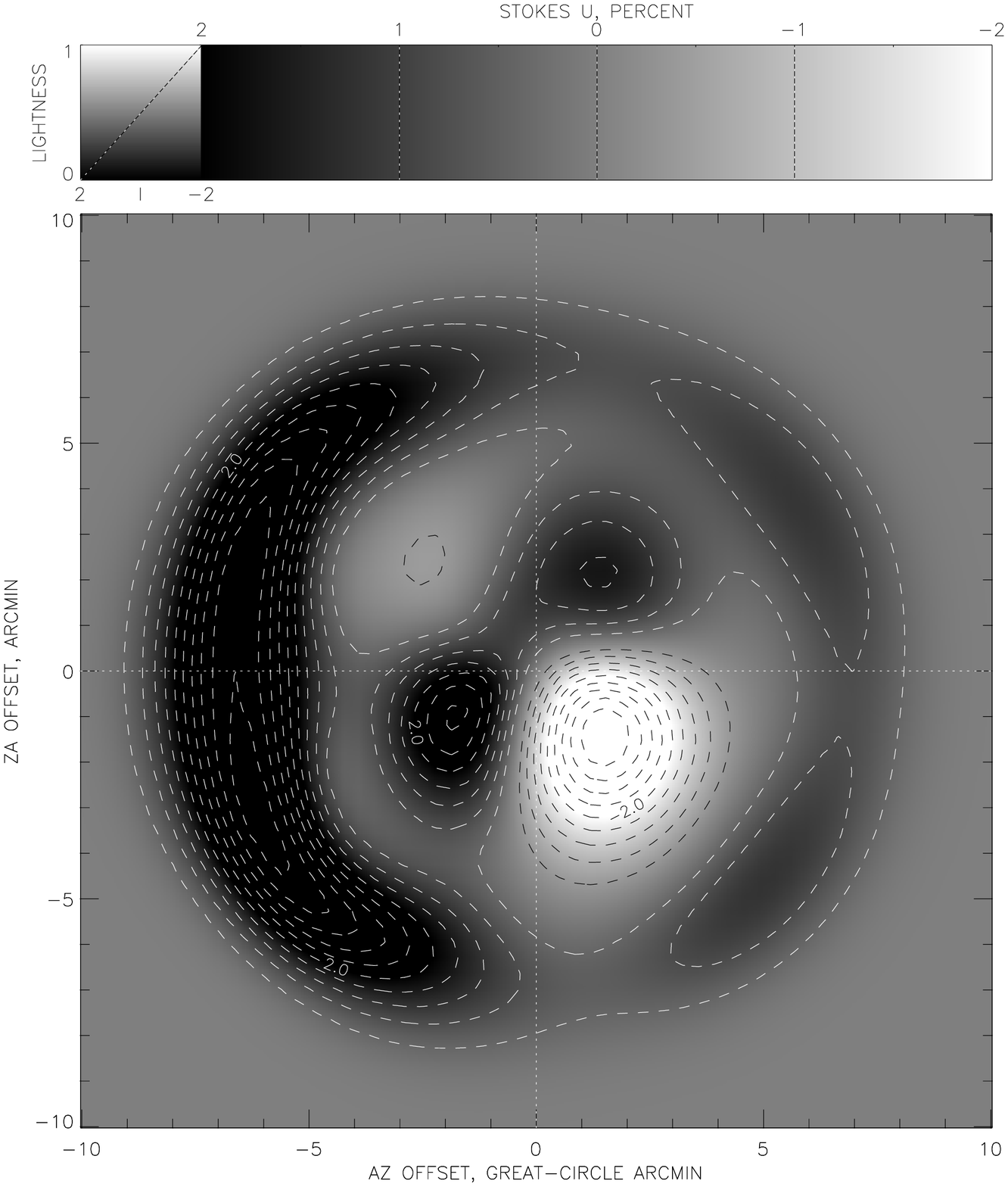}{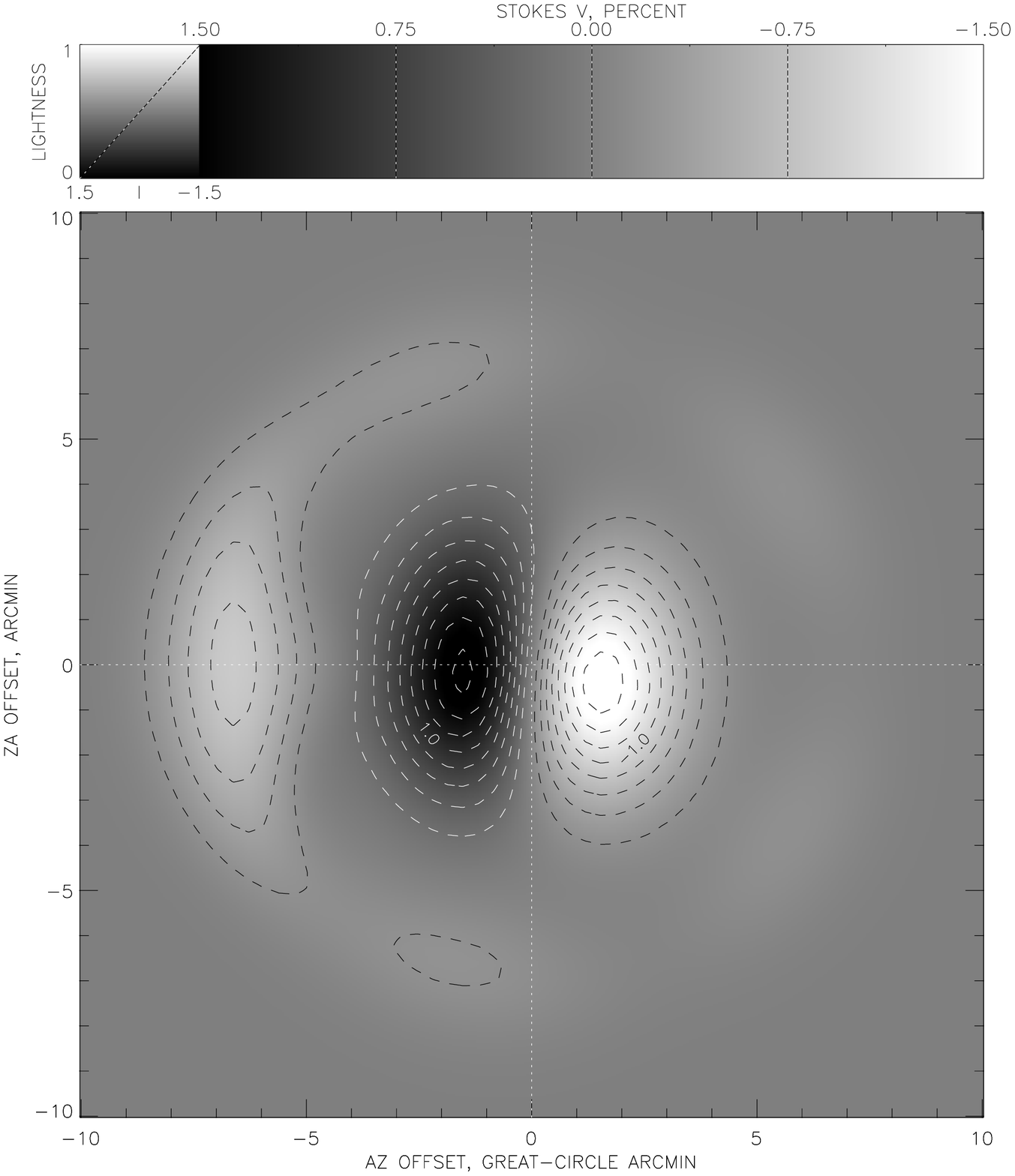}
\caption{Stokes $(I,Q,U,V)$ grey-scale/contour images of Arecibo's main
beam and first sidelobe near zenith angle $20\deg$.  For $I$, solid
(white) contours are $(0.1, 0.2, \dots)$ of the peak; dashed (black)
contours are $(0.01, 0.02, \dots)$. For the others,  black contours are
for areas with negative $(Q,U,V)$ with the grey scale tending towards 
white; white contours are positive $(Q,U,V)$ with the greyscale tending
towards black. Contours are in percent of Stokes $I$ at beam center and
spaced by $0.4\%$ for $(Q,U)$ and $0.2\%$ for $V$; the $0\%$ contours
are omitted. $U$ is aligned with the azimuth arm.  \label{mkfig1iquv} }
\end{figure}

\subsection{Main beam circular polarization}

	With circular polarization and a paraboloid having the feed
aligned perfectly at the focal point, everything is circularly symmetric
so there can be no beam structure in Stokes $V$. In practice, however,
such perfection can never be achieved. If the feed points slightly
away from the vertex of the paraboloid, say in the $X$ direction, then
the two circularly polarized beams point in slightly different
directions along the $Y$ direction. When the two circulars are
subtracted to produce Stokes $V$, there is a two-lobed structure in the
Stokes $V$ beam response. We call this {\it beam squint}; see Figure
\ref{mkfig1iquv}. 

	Arecibo and the GBT both employ an asymmetrically fed design in
which the feed is both located off-axis and doesn't point towards the
center of symmetry. Thus, both have intrinsic beam squint built into
their designs. One can minimize the polarization structure by
appropriate design of the subreflector geometry. This optimization was
done for both telescopes and, at least at Arecibo, the predicted main
beam Stokes $V$ performance is close to what's measured; the difference
in direction $\sim 2$ arcsec.

\subsection {Sidelobe polarization}

	Figure \ref{mkfig1iquv} shows the effects of beam
squash and squint. They also show that the first sidelobe is highly
polarized. As one moves away from beam center to encounter higher order
sidelobes, the effects of distortion increase and the sidelobe
polarizations increase. These effects are not well studied or
understood. 

	Arecibo has a large central blockage produced by the suspended
``triangle'' structure from which the moving feed hangs. This produces
large sidelobes (Heiles et al 2000b, 2001b), and these have high
polarization. The GBT, with its unblocked aperture, has exceedingly low
sidelobes, so effects arising from sidelobe polarization are minimized.

\subsection{ The effect on astronomical polarization measurements}

        Suppose one is observing a large-scale feature where the
brightness temperature $T_B$ varies with position.  One can express this
variation by a two-dimensional Taylor expansion. Beam squint, by its
nature, responds to the first derivative and only slightly to the
second; beam squash responds primarily to the second. 

	The polarized beam structure produces fake results in the {\it
polarized} Stokes parameters $(Q,U,V)$ that arise from spatial gradients
in the {\it total intensity} Stokes parameter $I$.   The effects are
exacerbated by the polarized sidelobes, which are further from beam
center.

	Heiles et al (2000b, 2001b) calculated these effects for the
Arecibo beam at 1.4 GHz, including both the main beam and first sidelobe
but no additional sidelobes. Consider a total intensity gradient of 1 K
arcmin$^{-1}$ and second derivative 1 K arcmin$^{-2}$, values which are
not necessarily realistic but are convenient for practical use. For
these particular values the  the fake contributions from the first and
second derivatives are comparable. They yield fake results for Stokes
${Q,U} \sim 0.3$ K. For Stokes $V$ the contributions are about ten times
smaller, $\sim 0.03$ K.

	The fractional polarization of extended emission tends to be
small, so spatial gradients in $I$ can be very serious.  Consider, for
example, measuring Zeeman splitting of the 21-cm line in emission, which
involves measuring Stokes $V$ of the 21-cm line. If the central velocity
of the 21-cm line has a spatial gradient ${dv \over d\theta} = 1$ km
s$^{-1}$ deg$^{-1}$---a not-uncommon value as measured with a 36 arcmin
beam (Heiles 1996)---we get $B_{fake} \sim 1.1$ $\mu$G.  Gradients might
be larger when measured with smaller $HPBW$.  Typical  values of $B$ are
in the $\mu$G range, so this effect can be---but is perhaps not
always---serious!

	In principle, these effects can be corrected for. Correcting for
them at Arecibo is a complicated business because of the $P\!A$
variation with azimuth and zenith angle.  It is also an uncertain
business, especially for $(Q,U)$ and somewhat less so for $V$, because
these variations are unpredictable and must be determined empirically.
Presumably, corrections at the GBT will be much more straightforward.

\section{Summary}

	After a brief introduction to the astrophysical significance of
polarization measurement, \S 2 began by introducing the Jones vector,
which describes the polarization state of a sine wave. \S 3 went on to 
define the Stokes parameters and the Stokes vector, which are required
to completely describe the polarization state of natural radiation,
which is always at least partly randomly polarized. \S3 also used the
Stokes parameters to define the conventionally used quantities,
fractional polarization and position angle, and cautioned against their
use in arithmetic operations. 

	\S 4 related the Jones vector to the Stokes vector; this
relationship tells exactly how radio astronomers measure all four Stokes
parameters simultaneously. However, the receiving system modifies the
incoming polarization with instrumental effects, which must be measured
and corrected for; \S 5 described the quantitative aspects of this
correction using Jones and Mueller matrices. It detailed the specific
cases of the amplifier chain and the mechanical rotation of the
telescope on the sky as instructive and most important examples. The
Mueller matrix for the amplifier chain leads naturally to a discussion
of the advantages of cross correlation for measuring small effects. It
also shows how one must not use this falsely, as is often done using a
post-amplifier hybrid. 

	Finally, \S 7 we described the major polarization effects in the
main beam, namely beam squint and beam squash, and illustrated these
effects using Arecibo as an example. Arecibo has a prominent first
sidelobe, and \S 7 also discussed its rather severe polarization
properties. It concluded by discussing of the effects of this beam
structure, interacting with spatial derivatives in Stokes $I$, on
contaminating measurements of polarized Stokes parameters.

\acknowledgements

	This work was supported in part by NSF grants 95-30590 and
AST-0097417.

\enddocument
\end